\documentclass[journal,draftclsnofoot,onecolumn,12pt,comsoc, twoside,a4paper]{IEEEtran}

\input{packages_commands_theorems.tex}

\usepackage[noamssymbols]{newtxmath}
\usepackage{eucal}

\usepackage[capitalise]{cleveref}

\usepackage{bm}

\newcommand{\diag}{\mathop{\mathrm{diag}}}

\newcommand{\stack}{\mathop{\mathrm{vec}}}

\newcommand*\diff{\mathop{}\!\mathrm{d}}

\usepackage{acronym}
\acrodef{CSI}[CSI]{channel state information}
\acrodef{CSI2}[CSI]{channel state information}
\acrodef{CSIT}[CSIT]{CSI at the transmitter}
\acrodef{RIS}[RIS]{reconfigurable intelligent surface}
\acrodef{RISs}[RISs]{Reconfigurable intelligent surfaces}
\acrodef{MMSE}[MMSE]{minimum mean-square error}
\acrodef{SNR}[SNR]{signal-to-noise ratio}
\acrodef{ASK}[ASK]{amplitude shift keying}
\acrodef{PSK}[PSK]{phase shift keying}
\acrodef{SCD}[SCD]{successive cancellation decoding}
\acrodef{MIMO}[MIMO]{multiple-input multiple-output}
\acrodef{PSK}[PSK]{phase-shift keying}
\acrodef{CGF}[CGF]{cumulant-generating function}
\acrodef{iid}[i.i.d.]{independent and identically distributed}
\acrodef{QAM}[QAM]{quadrature amplitude modulation}
\acrodef{RF}[RF]{radio frequency}
\acrodef{MAC}[MAC]{multiple access channel}
\acrodef{MU}[MU]{multi-user}

\begin{document}

\title{Single-RF Multi-User Communication Through
	Reconfigurable Intelligent Surfaces: An Information-Theoretic Analysis}

\author{Roy~Karasik,~\IEEEmembership{Student Member,~IEEE,}
	Osvaldo~Simeone,~\IEEEmembership{Fellow,~IEEE,}
	Marco~Di~Renzo,~\IEEEmembership{Fellow,~IEEE,}
	and~Shlomo~Shamai~(Shitz),~\IEEEmembership{Life Fellow,~IEEE}%
	\thanks{This work has been supported by the European Research Council (ERC) and by the Information and Communication Technologies (ICT) under the European Union’s Horizon 2020 Research and Innovation Programme (Grant Agreement Nos. 694630, 725731, and 871464). }
	\thanks{R. Karasik and S. Shamai are with the Department of Electrical Engineering, Technion --- Israel Institute of Technology, Haifa 32000, Israel. (royk@campus.technion.ac.il).}%
	\thanks{O. Simeone is with the Centre for Telecommunications Research,
		Department of Informatics, King’s College London, London WC2R 2LS, U.K.
		(osvaldo.simeone@kcl.ac.uk).}%
	\thanks{M. Di Renzo is with Universit\'e Paris-Saclay, CNRS, CentraleSup\'elec, Laboratoire des Signaux et Syst\`emes, 3 Rue Joliot-Curie, 91192 Gif-sur-Yvette, France.
		(marco.di-renzo@universite-paris-saclay.fr).}
}

\maketitle

\begin{abstract}
	\ac{RISs} are typically used in multi-user systems to mitigate interference among active transmitters. In contrast, this paper studies a setting with a conventional active encoder as well as a passive encoder that modulates the reflection pattern of the \acs{RIS}. The \acs{RIS} hence serves the dual purpose of improving the rate of the active encoder and of enabling communication from the second encoder. The capacity region is characterized, and information-theoretic insights regarding the trade-offs between the rates of the two encoders are derived by focusing on the high- and low-power regimes.
\end{abstract}

\section{Introduction}
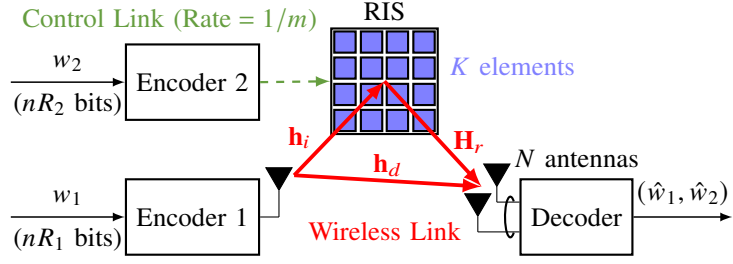
\begin{figure}[!t]
	\centering
	\begin{tikzpicture}[>=latex,
	ant/.style={%
		draw,
		fill,
		regular polygon,
		regular polygon sides=3,
		shape border rotate=180,
	},
]

\tikzset{
	partial ellipse/.style args={#1:#2:#3}{
		insert path={+ (#1:#3) arc (#1:#2:#3)}
	}
}

\node (EN) [thick,draw,minimum width=1cm,minimum height=1cm, font=\small] at (0,0) {Encoder 1};
\node (corner1) [right=0.1cm of EN] {};
\path[draw] (EN.east) -- (corner1.center);
\node (Ant1) [ant, above=0.2cm of corner1,scale=0.5] {};
\path[draw] (corner1.center) -- (Ant1);

\node (user) [thick,draw,minimum width=1cm,minimum height=1cm,right = 4.3cm of EN.center, font=\small]  {Decoder};
\node (RxP1) [left=0.15cm of user] {};
\node (RxP6) [left=0.40cm of user] {};
\node (RxP2) [above=0.05cm of RxP1.center] {};
\node (RxP3) [above=0.05cm of user.west] {};
\node (RxP4) [below=0.05cm of RxP6.center] {};
\node (RxP5) [below=0.05cm of user.west] {};
\path[draw] (RxP3.center) -- (RxP2.center);
\path[draw] (RxP5.center) -- (RxP4.center);
\node (Ant2) [ant, above =0.2cm of RxP2.center,scale=0.5] {};
\path[draw] (RxP2.center) -- (Ant2);
\node (Ant3) [ant, above =0.2cm of RxP4.center,scale=0.5] {};
\path[draw] (RxP4.center) -- (Ant3);
\node [font=\small, above=0.0mm of user.north,xshift=0mm] {$N$ antennas};
\draw[thick] ($(user.west)+(-1.2mm,0mm)$) [partial ellipse=50:310:0.7mm and 2.7mm];

\node (IRS) [thick,draw,minimum width=1.4cm,minimum height=1.4cm, font=\small] at ($(EN)!0.5!(user)+(0,1.8cm)$) {};
\node (R1) [thick,fill=blue!50,draw,minimum width=0.26cm,minimum height=0.26cm] at ([yshift=5.2mm,xshift=-5.2mm]IRS.center){};
\node (R2) [thick,fill=blue!50,draw,minimum width=0.26cm,minimum height=0.26cm] at ([yshift=-5.2mm,xshift=-5.2mm]IRS.center){};
\node (R3) [thick,fill=blue!50,draw,minimum width=0.26cm,minimum height=0.26cm] at ([yshift=5.2mm,xshift=+5.2mm]IRS.center){};
\node (R4) [thick,fill=blue!50,draw,minimum width=0.26cm,minimum height=0.26cm] at ([yshift=-5.2mm,xshift=+5.2mm]IRS.center){};
\node (R5) [thick,fill=blue!50,draw,minimum width=0.26cm,minimum height=0.26cm] at ([yshift=+5.2mm,xshift=+1.7mm]IRS.center){};
\node (R6) [thick,fill=blue!50,draw,minimum width=0.26cm,minimum height=0.26cm] at ([yshift=+5.2mm,xshift=-1.7mm]IRS.center){};
\node (R7) [thick,fill=blue!50,draw,minimum width=0.26cm,minimum height=0.26cm] at ([yshift=-5.2mm,xshift=+1.7mm]IRS.center){};
\node (R8) [thick,fill=blue!50,draw,minimum width=0.26cm,minimum height=0.26cm] at ([yshift=-5.2mm,xshift=-1.7mm]IRS.center){};
\node (R9) [thick,fill=blue!50,draw,minimum width=0.26cm,minimum height=0.26cm] at ([yshift=1.75mm,xshift=-5.2mm]IRS.center){};
\node (R10) [thick,fill=blue!50,draw,minimum width=0.26cm,minimum height=0.26cm] at ([yshift=-1.75mm,xshift=-5.2mm]IRS.center){};
\node (R11) [thick,fill=blue!50,draw,minimum width=0.26cm,minimum height=0.26cm] at ([yshift=1.75mm,xshift=+5.2mm]IRS.center){};
\node (R12) [thick,fill=blue!50,draw,minimum width=0.26cm,minimum height=0.26cm] at ([yshift=-1.75mm,xshift=+5.2mm]IRS.center){};
\node (R13) [thick,fill=blue!50,draw,minimum width=0.26cm,minimum height=0.26cm] at ([yshift=+1.75mm,xshift=+1.7mm]IRS.center){};
\node (R14) [thick,fill=blue!50,draw,minimum width=0.26cm,minimum height=0.26cm] at ([yshift=+1.75mm,xshift=-1.7mm]IRS.center){};
\node (R15) [thick,fill=blue!50,draw,minimum width=0.26cm,minimum height=0.26cm] at ([yshift=-1.75mm,xshift=+1.7mm]IRS.center){};
\node (R16) [thick,fill=blue!50,draw,minimum width=0.26cm,minimum height=0.26cm] at ([yshift=-1.75mm,xshift=-1.7mm]IRS.center){};
\node [font=\small, above=-0.5mm of IRS.north] {RIS};
\node [font=\small, right=0.0mm of IRS.east,yshift=2mm,blue!50] {$K$ elements};

\node (EN2P1) at ($(IRS.west)+(0,0mm)$) {};
\node (EN2) [thick,draw,minimum width=1cm,minimum height=1cm, font=\small, above = 0.76cm of EN] {Encoder 2};

\coordinate (A) at ($(Ant1)+(2mm,0)$);
\path[draw,->,line width=0.5mm,red] (A) -- node[above,font=\small,pos=0.15,xshift=-1mm] {$\mathbf{h}_i$} coordinate[pos=0.2] (P1) coordinate[pos=0.25] (P2) ($(IRS)+(0,0)$);

\coordinate (B) at ($(Ant2)+(-2mm,-2mm)$);
\path[draw,->,line width=0.5mm,red] ($(IRS)+(0,0)$) -- node[above,font=\small,pos=0.8,xshift=1mm] {$\mathbf{H}_{r}$} coordinate[pos=0.75] (P3) coordinate[pos=0.8] (P4) (B);

\path[draw,->,line width=0.5mm,red] (A) -- node[above,font=\small,pos=0.5,yshift=-1mm] {$\mathbf{h}_d$} coordinate[pos=0.75] (P5) coordinate[pos=0.78] (P6) (B);

\node [font=\small, below=10mm of IRS,red] {Wireless Link};

\path[draw,->,line width=0.2mm] ($(EN.west)-(15mm,0)$) -- node[above,font=\small] {$w_1$} node[below,font=\small] {($nR_1$ bits)} ($(EN.west)+(0,0)$);
\path[draw,->,line width=0.2mm] ($(EN2.west)-(15mm,0)$) -- node[above,font=\small] {$w_2$} node[below,font=\small] {($nR_2$ bits)} ($(EN2.west)+(0,0)$);
\path[draw,->,line width=0.2mm] ($(user.east)-(0mm,0)$) -- node[above,font=\small] {$(\hat{w}_1,\hat{w}_2)$} ($(user.east)+(13mm,0)$);

\draw[line width=0.3mm,OliveGreen,->,dashed] (EN2.east) -- (EN2P1.center);
\node [font=\small, above=0.0mm of EN2.north, OliveGreen, xshift=-3mm] {Control Link ($\text{Rate}=1/m$)};

\end{tikzpicture}
	\caption{In the system under study, Encoder 1 is active and it encodes its message $w_1$ into a codeword of $n$ symbols sent on the wireless link; whereas Encoder 2 is passive and it encodes the message $w_2$ into a control action, which is sent on the control link to the RIS at a rate of one action every $m$ channel symbols.}
	\label{fig:simple-model}
\end{figure}

A \acf{RIS} is a nearly-passive device that can shape the wireless propagation channel by applying phase shifts to the incident signals \cite{renzo2020Smart,renzo202Reconfigurable,wu2020towards,liu2020reconfigurable,renzo2019smart,yuan2020reconfigurable,wu2020intelligent}. 
In \ac{MU} systems, \ac{RIS}s can help mitigate inter-user interference and obtain beamforming gain for standard active transmitters \cite{guo2019weighted2,zhang2020reconfigurable,abrardo2020intelligent,mu2020joint,huang2019reconfigurable,han2020intelligent,zhang2020physical,zhou2020fairness}. To this end, the configuration of the \ac{RIS} is kept fixed for the duration of a coherence interval and optimized to maximize some function of the achievable rates \cite{guo2019weighted2,zhang2020reconfigurable,abrardo2020intelligent,mu2020joint,huang2019reconfigurable,han2020intelligent,zhang2020physical,zhou2020fairness}. In this paper, we study a different use of \ac{RIS}s, whereby a single active transmitter coexists with a \emph{passive} user, having no direct \ac{RF} chain, that conveys its own message by modulating the reflection pattern of the \ac{RIS} (see \cref{fig:simple-model}). 
With reference to \cref{fig:simple-model}, the \ac{RIS} is accordingly used for the dual purpose of enhancing the rate of the active encoder (Encoder 1) and of enabling communication for the passive encoder (Encoder 2). Unlike prior work \cite{yang2020novel} that focused on a specific transmission strategy, this paper concentrates on the information-theoretic analysis of the rate trade-offs between the two encoders, providing fundamental insights.

\emph{Related Work:} A comprehensive survey of the state-of-the-art on \ac{RIS}-aided \ac{MU} systems is available in \cite{renzo2020Smart}. As notable representative examples of works involving active transmitters, the maximization of the weighted sum-rate in \ac{RIS}-aided \ac{MU} systems was studied in \cite{guo2019weighted2,zhang2020reconfigurable,abrardo2020intelligent,mu2020joint}, whereas references \cite{huang2019reconfigurable,han2020intelligent} focused on optimizing the energy efficiency, and papers \cite{zhang2020physical,zhou2020fairness} on physical-layer security and outage-probability enhancements. A \ac{MU} system with an active transmitter and a passive encoder, akin to \cref{fig:simple-model}, was proposed in \cite{yang2020novel} by assuming binary modulation, a single receiver antenna, and a specific successive interference cancellation decoding strategy. 

From an information-theoretic perspective, the single-\ac{RF} \ac{MU} communication system depicted in \cref{fig:simple-model} can be viewed as a \ac{MAC} with both multiplicative and additive elements. The capacity of the Gaussian multiplicative \ac{MAC} was derived in \cite{pillai2011on} for two active encoders. The capacity region of a backscatter multiplicative \ac{MAC}, which can be viewed as a special case of the \ac{RIS}-aided \ac{MU} communication system in \cref{fig:simple-model} with one reflecting element, was studied in \cite{liu2018backscatter}. Under the assumptions of a single receiver antenna and Gaussian codebooks, this work shows that conventional time-sharing schemes are suboptimal in the high-power and weak-backscatter regimes. The capacity of an \ac{RIS}-aided \emph{single-user} channel was derived in \cite{karasik2020adaptive}.

\emph{Main Contributions:} In this paper, we study the \ac{RIS}-aided \ac{MU} system illustrated in \cref{fig:simple-model}, in which Encoder 1 is active, whereas Encoder 2 can only alter the reflection pattern of the \ac{RIS} in a passive manner. We derive the capacity region under the practical assumptions of a multi-antenna decoder, a finite-input constellation, and a set of discrete phase shifts at the \ac{RIS}. Then, we specialize the results for the high- and low-power regimes, showing that: \emph{(i)} for sufficiently high transmission power, both encoders can simultaneously communicate at maximum rate; and \emph{(ii)} in the low-power regime, Encoder 1 can achieve maximum rate if and only if Encoder 2 does not communicate, while Encoder 2 can achieve its maximum rate while still enabling non-zero-rate communication for Encoder 1. Finally, numerical examples demonstrate the dual role of the \ac{RIS} as means to enhance the transmitted signal on the one hand and as the enabler of \ac{MU} communication on the other hand.

\emph{Notation:} 
Random variables, vectors, and matrices are denoted by lowercase, boldface lowercase, and boldface uppercase Roman-font letters, respectively. Realizations of random variables, vectors, and matrices are denoted by lowercase, boldface lowercase, and boldface uppercase italic-font letters, respectively. For example, $x$ is a realization of random variable $\mathrm{x}$, $\bm{x}$ is a realization of random vector $\mathbf{x}$, and $\bm{X}$ is a realization of random matrix $\mathbf{X}$.
For any positive integer $K$, we define the set $[K]\triangleq \{1,2,\ldots,K\}$. 
The cardinality of a set $\mathcal A$ is denoted as $|\mathcal{A}|$. 
The $\ell^2$-norm and the conjugate transpose of a vector $\bm{v}$ are denoted as $\lVert\bm{v}\rVert$ and $\bm{v}^\ast$, respectively.
$\diag(\bm{x})$ represents a diagonal matrix with diagonal given by the vector $\bm{x}$. 
The vectorization of matrix $\bm{H}$, i.e., the operator that stacks the columns of $\bm{H}$ on top of one another, is denoted by $\stack(\bm{H})$. 
The Kronecker product $\bm{I}_m\otimes\bm{B}$ of the identity matrix of size $m$ and matrix $\bm{B}$  is denoted as $\bm{B}^{m\otimes}$.

\section{System Model}\label{sec:model}
We consider the system depicted in \cref{fig:simple-model}
in which two encoders communicate with a decoder equipped with $N$ antennas over a quasi-static fading channel in the presence of an \ac{RIS} that comprises $K$ nearly-passive reconfigurable elements. Encoder 1 is equipped with a single-RF transmitter and encodes its message $w_1\in[2^{nR_1}]$ of rate $R_1$ [bits/symbol] into a codeword of $n$ symbols sent on the wireless link to the decoder. In contrast, Encoder 2 encodes its message $w_2\in[2^{nR_2}]$ of rate $R_2$ [bits/symbol] in a passive manner by modulating the reflection pattern of the \ac{RIS}. The reflection pattern is controlled through a rate-limited control link, and is defined by the phase shifts that each of the $K$ \ac{RIS} elements applies to the impinging wireless signal. Encoder 2 represents, for example, a sensor embedded in the \ac{RIS} that applies metasurface-based modulation in order to convey its sensed data without emitting a new radio wave \cite[Sec. 3.3]{renzo2019smart}.

A coding slot consists of $n$ symbols, which are divided into $n/m$ blocks of $m$ symbols each, with $n/m$ assumed to be integer. Specifically, the codeword transmitted by Encoder 1 as a function of message $w_1$ occupies the entire coding slot, and it includes $n$ symbols from a constellation $\mathcal{S}$ of $S=|\mathcal S|$ points. Furthermore, the \ac{RIS} is controlled by Encoder 2 by selecting the phase shift applied by each of the $K$ elements of the \ac{RIS} from a finite set $\mathcal{A}$ of $A=|\mathcal{A}|$ distinct hardware-determined values as a function of the message $w_2$.
Due to practical limitations on the \ac{RIS} control rate, the phase shifts can only be modified once for each block of $m$ consecutive transmitted symbols. 
During the $t$th block, the fraction of the codeword of Encoder 1 consisting of $m$ transmitted symbols is denoted by $\mathbf{s}(t)=(\mathrm{s}_{1}(t),\ldots,\mathrm{s}_{m}(t))^\intercal\in\mathcal S^{m\times 1}$, and is assumed to satisfy the power constraint 
\begin{IEEEeqnarray}{c}\label{eq:power_constraint}
	\frac{1}{m}\mathbb E[\mathbf{s}^\ast(t)\mathbf{s}(t)]\leq 1.
\end{IEEEeqnarray}
The phase shifts applied by the \ac{RIS} in the $t$th block are denoted by the vector
\begin{IEEEeqnarray}{c}
	e^{j\pmb{\uptheta}(t)}\triangleq (e^{j\uptheta_{1}(t)},\ldots,e^{j\uptheta_{K}(t)})^\intercal,
\end{IEEEeqnarray}
with $\uptheta_{k}(t)\in\mathcal{A}$ being the phase shift applied by the $k$th \ac{RIS} element, $k\in[K]$.

We assume quasi-static flat-fading wireless channels, which remain fixed throughout a coding slot. Specifically, the channel from Encoder 1 to the decoder is denoted by vector $\mathbf{h}_d\in\mathbb C^{M\times 1}$; the channel from Encoder 1 to the \ac{RIS} is denoted by the vector $\mathbf{h}_i\in\mathbb C^{K\times 1}$; and the channel from the \ac{RIS} to the $N$ receiving antennas is denoted by the matrix $\mathbf{H}_r\in\mathbb C^{N\times K}$. Furthermore, we assume that $\mathbf{h}_d$, $\mathbf{h}_i$, and $\mathbf{H}_r$ are drawn from a continuous distribution. Finally, we denote the signal received by the $N$ antennas for the $q$th transmitted symbol in block $t\in[n/m]$ by $\mathbf{y}_{q}(t)\in\mathbb C^{N\times 1}$, $q\in[m]$. The overall received signal matrix $\mathbf{Y}(t)=(\mathbf{y}_{1}(t),\ldots,\mathbf{y}_{m}(t))\in\mathbb C^{N\times m}$ in the $t$th sub-block can hence be written as
\begin{IEEEeqnarray}{rCl}\label{eq:channel}
	\mathbf{Y}(t)&=&\sqrt{P}\mathbf{H}_r\diag\left(e^{j\pmb{\uptheta}(t)}\right)\mathbf{h}_i\mathbf{s}^\intercal(t)+\mathbf{h}_d\mathbf{s}^\intercal(t)+\mathbf{Z}(t)\IEEEnonumber\\
	&=&\sqrt{P}\rb{\mathbf{H}_{ri}e^{j\pmb{\uptheta}(t)}+\mathbf{h}_d}\mathbf{s}^\intercal(t)+\mathbf{Z}(t),
\end{IEEEeqnarray}
where $P>0$ denotes the transmission power of Encoder 1; the matrix $\mathbf{H}_{ri}\triangleq \mathbf{H}_r\diag(\mathbf{h}_i)\in\mathbb{C}^{N\times K}$, combines the channels $\mathbf{h}_i$ and $\mathbf{H}_r$; 
and the matrix $\mathbf{Z}(t)\in\mathbb C^{N\times m}$, whose elements are \ac{iid} as $\mathcal{CN}(0,1)$, denotes the additive white Gaussian noise at the receiving antennas.
In order to characterize the distribution of the output signal $\mathbf{Y}(t)$ in \eqref{eq:channel}, we vectorize it as
\begin{IEEEeqnarray}{c}\label{eq:vec_channel}
	\mathbf{y}(t)\triangleq \stack(\mathbf{Y}(t))=\sqrt{P}\rb{\mathbf{H}_{ri}e^{j\pmb{\uptheta}(t)}+\mathbf{h}_d}^{m\otimes}\mathbf{s}(t)+\mathbf{z}(t),
\end{IEEEeqnarray}
where we have defined the vector $\mathbf{z}(t)\triangleq\stack(\mathbf{Z}(t))\in\mathbb{C}^{Nm\times 1}$.

We assume that both the encoders and the decoder have perfect \ac{CSI}, in the sense that the channel matrix $\mathbf{H}_{ri}$ and channel vector $\mathbf{h}_d$ are known.
Having received signal $\mathbf{y}(t)$ in \eqref{eq:vec_channel} for $t\in[n/m]$, the decoder produces the estimates $\hat{w}_\ell=\hat{w}_\ell(\mathbf{y}(1),\ldots,\mathbf{y}(n/m),\mathbf{H}_{ri},\mathbf{h}_d)$, for $\ell=1,2$, using knowledge of the \ac{CSI}.
Given channel realizations $\bm{H}_{ri}$ and $\bm{h}_d$, a rate pair $\rb{R_1(\bm{H}_{ri},\bm{h}_d),R_2(\bm{H}_{ri},\bm{h}_d)}$ is said to be \emph{achievable} if the probability of error satisfies the limit $\Pr(\hat{w}_1\neq w_1,\hat{w}_2\neq w_2)\rightarrow 0$ when the codeword length grows large, i.e., $n\rightarrow\infty$. The corresponding capacity region $\mathcal{C}(\bm{H}_{ri},\bm{h}_d)$ is defined as the closure of the set of achievable rate pairs. 

\section{Capacity Region}\label{sec:capacity_region}
In this section, we first derive a general characterization of the capacity region $\mathcal{C}(\bm{H}_{ri},\bm{h}_d)$ for the channel in \eqref{eq:vec_channel}. Then, we leverage this result to provide theoretical insights into the trade-offs between the rate of the two encoders in \cref{fig:simple-model} by focusing on the low- and high-power regimes.

Most existing works on the multiplicative Gaussian \ac{MAC} \cite{pillai2011on,liu2018backscatter} and on \ac{RIS}-aided systems (see, e.g., \cite{guo2019weighted2}) consider Gaussian codebooks for the transmitted signal $\mathbf{s}(t)$. This implies that the resulting achievable rates are formulated in the standard form ``$\log_2(1+\text{SNR})$''. In contrast, as described in \cref{sec:model}, we focus our attention on the more practical model in which the transmitted symbols and the \ac{RIS} elements' phase response take values from finite sets \cite{wu2018survey}. Therefore, in a manner similar to \cite{karasik2020adaptive}, the expressions for the achievable rates that we present in this section are more complex, and require the following definition. 

\begin{definition}
	The \ac{CGF} of a random variable $\mathrm{u}$ conditioned on a random vector $\mathbf{x}$ is defined as
	\begin{IEEEeqnarray}{c}\label{eq:def_cond_cgf_rand}
		\kappa_r(\mathrm{u}|\mathbf{x})\triangleq \mathbb E\sqb{\log_2\rb{\mathbb E\sqb{e^{r\mathrm{u}}|\mathbf{x}}}},\quad \text{for }r\in\mathbb R,
	\end{IEEEeqnarray}
	and the value of the conditional \ac{CGF} for $r=1$ is denoted as $\kappa(\mathrm{u}|\mathbf{x})\triangleq\kappa_1(\mathrm{u}|\mathbf{x})$.
\end{definition}

We now characterize the capacity region in the form of a union of rate regions, with each region corresponding to rates achievable for a specific choice of encoding distributions $p_{\mathbf{s}}(\bm{s})$ and $p_{\pmb{\uptheta}}(\pmb{\theta})$ of the transmitted symbols $\mathbf{s}(t)$ and \ac{RIS} phase shifts $\pmb{\uptheta}(t)$ in \eqref{eq:vec_channel}, respectively. 

\begin{proposition}\label{prop:capacity_region}
	For input distributions $p_{\mathbf{s}}(\bm{s})$ and $p_{\pmb{\uptheta}}(\pmb{\theta})$, let 
	$\mathcal{R}(p_{\mathbf{s}},p_{\pmb{\uptheta}},\bm{H}_{ri},\bm{h}_d)$ be the set of rate pairs $\rb{R_1(\bm{H}_{ri},\bm{h}_d),R_2(\bm{H}_{ri},\bm{h}_d)}$ such that the inequalities
	\begin{IEEEeqnarray}{c}\label{eq:capacity_region}
		R_\ell(\bm{H}_{ri},\bm{h}_d)\leq -N\log_2(e)-\frac{1}{m}\kappa(\mathrm{u}_\ell|\mathbf{s}_1,\pmb{\uptheta}_1,\mathbf{z}),\quad \ell\in\{1,2\},\IEEEyesnumber\IEEEyessubnumber
	\end{IEEEeqnarray}
	and
	\begin{IEEEeqnarray}{c}
		\!\! R_1(\bm{H}_{ri},\bm{h}_d)+R_2(\bm{H}_{ri},\bm{h}_d)\leq-N\log_2(e)-\frac{1}{m}\kappa(\mathrm{u}_3|\mathbf{s}_1,\pmb{\uptheta}_1,\mathbf{z})\IEEEyessubnumber
	\end{IEEEeqnarray}
	hold, where random variable $\mathrm{u}_1$, $\mathrm{u}_2$, and $\mathrm{u}_3$ are defined as
	\begin{IEEEeqnarray}{rCl}
		\mathrm{u}_1&\triangleq& -\Big\lVert\mathbf{z}+\sqrt{P}\rb{\bm{H}_{ri}e^{j\pmb{\uptheta}_1}+\bm{h}_d}^{m\otimes}\rb{\mathbf{s}_1-\mathbf{s}_2}\Big\rVert^2,\IEEEyesnumber\IEEEyessubnumber\\
		\mathrm{u}_2&\triangleq& -\Big\lVert\mathbf{z}+\sqrt{P}\rb{\bm{H}_{ri}\sqb{e^{j\pmb{\uptheta}_1}-e^{j\pmb{\uptheta}_2}}}^{m\otimes}\mathbf{s}_1\Big\rVert^2,\IEEEyessubnumber\\
		\mathrm{u}_3&\triangleq& -\Big\lVert\mathbf{z}+\sqrt{P}\rb{\bm{H}_{ri}e^{j\pmb{\uptheta}_1}+\bm{h}_d}^{m\otimes}\mathbf{s}_1
		-\sqrt{P}\rb{\bm{H}_{ri}e^{j\pmb{\uptheta}_2}+\bm{h}_d}^{m\otimes}\mathbf{s}_2\Big\rVert^2,\IEEEyessubnumber
	\end{IEEEeqnarray}
	respectively, with independent random vectors $\mathbf{s}_1,\mathbf{s}_2\sim p_{\mathbf{s}}(\bm{s})$, $\pmb{\uptheta}_1,\pmb{\uptheta}_2\sim p_{\pmb{\uptheta}}(\pmb{\theta})$, and $\mathbf{z}\sim\mathcal{CN}(\mathbf{0},\bm{I}_{Nm})$.
	The capacity region $\mathcal{C}(\bm{H}_{ri},\bm{h}_d)$ is the convex hull of the union of the regions $\mathcal{R}(p_{\mathbf{s}},p_{\pmb{\uptheta}},\bm{H}_{ri},\bm{h}_d)$ over all input distributions $p_{\mathbf{s}}(\bm{s})$ and $p_{\pmb{\uptheta}}(\pmb{\theta})$ with $\bm{s}\in\mathcal{S}^{m\times 1}$, $\pmb{\theta}\in\mathcal{A}^{K\times 1}$, such that $\mathbb E[\mathbf{s}^\ast\mathbf{s}]\leq m$.
\end{proposition}
\begin{IEEEproof}
	See Appendix \ref{app:proof_capacity_region}.
\end{IEEEproof}

Next, we specialize the results in \cref{prop:capacity_region} to characterize the capacity region in the high- and low-power regimes.

\subsection{High-Power Regime}
The following corollary shows that the capacity region $\mathcal{C}(\bm{H}_{ri},\bm{h}_d)$ converges to a rectangle as the power of Encoder 1 increases.
\begin{corollary}\label{cor:high_snr}
	For any finite constellation $\mathcal S$ of $S=|\mathcal S|$ points and any set $\mathcal{A}$ of $A=|\mathcal A|$ phases, let $\overline{\mathcal C}$ be the set of rate pairs $\rb{R_1,R_2}$ such that
	\begin{IEEEeqnarray}{c}\label{eq:capacity_region_high_power}
		\overline{\mathcal C}\triangleq\cb{\rb{R_1,R_2}:R_1\leq \log_2(S),~ R_2\leq \frac{K}{m}\log_2(A)}.
	\end{IEEEeqnarray}
	The capacity region $\mathcal{C}(\bm{H}_{ri},\bm{h}_d)$ converges to $\overline{\mathcal C}$ as the power $P$ increases in the sense that $\mathcal{C}(\bm{H}_{ri},\bm{h}_d)\subseteq \overline{\mathcal C}$, and there exists a sequence of achievable rate pairs $\rb{R_1(\bm{H}_{ri},\bm{h}_d),R_2(\bm{H}_{ri},\bm{h}_d)}\in \mathcal{C}(\bm{H}_{ri},\bm{h}_d)$ such that, almost surely,
	\begin{IEEEeqnarray}{rCl}\label{eq:high_snr_limits}
		\lim_{P\rightarrow\infty}R_1(\bm{H}_{ri},\bm{h}_d)&=&\log_2(S),\IEEEyesnumber\IEEEyessubnumber\\
		\lim_{P\rightarrow\infty}R_2(\bm{H}_{ri},\bm{h}_d)&=&\frac{K}{m}\log_2(A)\IEEEyessubnumber.
	\end{IEEEeqnarray}
\end{corollary}
\begin{IEEEproof}
	See Appendix \ref{app:proof_high_snr}.
\end{IEEEproof}

\cref{cor:high_snr} implies that, for sufficiently high power $P$, both encoders can simultaneously achieve their maximum rates. As a result, while not useful in increasing the high-power rate of Encoder 1, the presence of the \ac{RIS} enables communication at the maximum rate for Encoder 2 without creating deleterious interference on Encoder 1's transmission. 

\subsection{Low-Power Regime}
In this section, we characterize the capacity region $\mathcal{C}(\bm{H}_{ri},\bm{h}_d)$ in the low-power regime. To simplify the analysis, we focus on a system with one receiver antenna, $N=1$, and an \ac{RIS} control ratio of $m=1$. For this special case, the channel \eqref{eq:vec_channel} can be written as
\begin{IEEEeqnarray}{c}
	\mathrm{y}(t)=\sqrt{P}\rb{\mathbf{h}_{ri}^\intercal e^{j{\pmb{\uptheta}}(t)}+\mathrm{h}_d}\mathrm{s}(t)+\mathrm{z}(t),
\end{IEEEeqnarray}
where $\mathbf{h}_{ri}\in\mathbb{C}^{K\times 1}$ and $\mathrm{h}_d\in\mathbb{C}$ denote the reflected and direct channel paths, respectively, and $\mathrm{z}(t)\sim\mathcal{CN}(0,1)$ denotes the additive white Gaussian noise. Furthermore, we assume that the phase shift applied by each element of the \ac{RIS} is chosen from a finite set of $A$ uniformly spaced phases, i.e., $\mathcal A=\{0,2\pi/A,\ldots,2\pi(A-1)/A\}$; and that $\mathcal{S}$ is a zero-mean input constellation, i.e., 
\begin{IEEEeqnarray}{c}
	\sum_{s\in\mathcal{S}}s=0,
\end{IEEEeqnarray}
which is known to achieve the minimum energy per bit in many single-user channels \cite{verdu2002spectral,massey1976all,lapidoth2002fading}.

In order to formulate the capacity region in the low-power regime, we define the normalized rate $r_\ell(\bm{h}_{ri},h_d)$, $\ell\in\{1,2\}$, for unit of power as
\begin{IEEEeqnarray}{c}
	r_\ell(\bm{h}_{ri},h_d)\triangleq \lim_{P\rightarrow 0}\frac{R_\ell(\bm{h}_{ri},h_d)}{P}.
\end{IEEEeqnarray}
The capacity region in the low-power regime $\underline{\mathcal{C}}(\bm{h}_{ri},h_d)$ is accordingly defined as the closure of the set of achievable normalized rate pairs (see, e.g., \cite{caire2004suboptimality}).
\begin{proposition}\label{prop:low_snr}
	For input distributions $p_{\mathrm{s}}(s)$ and $p_{\pmb{\uptheta}}(\pmb{\theta})$, let 
	$\underline{\mathcal{R}}(p_{\mathrm{s}},p_{\pmb{\uptheta}},\bm{h}_{ri},h_d)$ be the set of normalized rate pairs $\rb{r_1(\bm{h}_{ri},h_d),r_2(\bm{h}_{ri},h_d)}$ such that the inequalities
	\begin{IEEEeqnarray}{l}\label{eq:capacity_region_low_power}
		r_\ell(\bm{h}_{ri},h_d)\leq \frac{\mathbb E[\underline{\mathrm{u}}_\ell]}{\ln(2)},\quad \ell\in\{1,2\},\IEEEyesnumber\IEEEyessubnumber\\
		\text{and }r_1(\bm{h}_{ri},h_d)+r_2(\bm{h}_{ri},h_d)\leq\frac{\mathbb E[\underline{\mathrm{u}}_3]}{\ln(2)}\IEEEyessubnumber
	\end{IEEEeqnarray}
	hold, where random variable $\underline{\mathrm{u}}_1$, $\underline{\mathrm{u}}_2$, and $\underline{\mathrm{u}}_3$ are defined as
	\begin{IEEEeqnarray}{rCl}\label{eq:def_u_low_power}
		\underline{\mathrm{u}}_1&\triangleq& \Big\lvert\rb{\bm{h}_{ri}^\intercal e^{j\pmb{\uptheta}_1}+h_d}\rb{\mathrm{s}_1-\mathrm{s}_2}\Big\rvert^2,\IEEEyesnumber\IEEEyessubnumber\\
		\underline{\mathrm{u}}_2&\triangleq& \Big\lvert\bm{h}_{ri}^\intercal\rb{e^{j\pmb{\uptheta}_1}-e^{j\pmb{\uptheta}_2}}\mathrm{s}_1\Big\rvert^2,\IEEEyessubnumber\\
		\underline{\mathrm{u}}_3&\triangleq& \Big\lvert\rb{\bm{h}_{ri}^\intercal e^{j\pmb{\uptheta}_1}+h_d}\mathrm{s}_1-\rb{\bm{h}_{ri}^\intercal e^{j\pmb{\uptheta}_2}+h_d}\mathrm{s}_2\Big\rvert^2,\IEEEyessubnumber
	\end{IEEEeqnarray}
	respectively, with independent random variables $\mathrm{s}_1,\mathrm{s}_2\sim p_{\mathrm{s}}(s)$ and random vectors $\pmb{\uptheta}_1,\pmb{\uptheta}_2\sim p_{\pmb{\uptheta}}(\pmb{\theta})$.
	The capacity region in the low-power regime $\underline{\mathcal{C}}(\bm{h}_{ri},h_d)$ is the convex hull of the union of the regions $\underline{\mathcal{R}}(p_{\mathrm{s}},p_{\pmb{\uptheta}},\bm{h}_{ri},h_d)$ over all input distributions $p_{\mathrm{s}}(s)$ and $p_{\pmb{\uptheta}}(\pmb{\theta})$ with $s\in\mathcal{S}$, $\pmb{\theta}\in\mathcal{A}^{K\times 1}$, such that $\mathbb E[|\mathrm{s}|^2]\leq 1$.
\end{proposition}
\begin{IEEEproof}
	See Appendix \ref{app:proof_low_snr}. 
\end{IEEEproof}

Unlike the high-power regime, the low-power capacity region \eqref{eq:capacity_region_low_power} is not a rectangle, implying that it is not possible for both encoders to communicate at their respective maximum rates. The next corollary elaborates on this point.
\begin{corollary}\label{cor:low_power_corner_points}
	Let $\tilde{\pmb{\theta}}$ be the beamforming phase-shift vector maximizing Encoder 1's rate, i.e.,
	\begin{IEEEeqnarray}{c}\label{eq:beamforming}
		\tilde{\pmb{\theta}}\triangleq \argmax_{\pmb{\theta}\in\mathcal{A}^{K\times 1}}\Big\lvert\bm{h}_{ri}^\intercal e^{j\pmb{\theta}}+h_d\Big\rvert^2.
	\end{IEEEeqnarray}
	In the low-power regime, Encoder 1 can achieve its maximum normalized rate
	\begin{IEEEeqnarray}{c}
		r_1(\bm{h}_{ri},h_d)=\frac{2}{\ln(2)}\Big\lvert\bm{h}_{ri}^\intercal e^{j\tilde{\pmb{\theta}}}+h_d\Big\rvert^2
	\end{IEEEeqnarray}
	if and only if Encoder 2 does not communicate, i.e., $r_2(\bm{h}_{ri},h_d)=0$. In contrast, if $\lvert\bm{h}_{ri}^\intercal e^{j\tilde{\pmb{\theta}}}+h_d\rvert^2>\lVert \bm{h}_{ri}\rVert^2$, Encoder 2 can achieve its maximum normalized rate
	\begin{IEEEeqnarray}{c}
		r_2(\bm{h}_{ri},h_d)=\frac{2}{\ln(2)}\lVert \bm{h}_{ri}\rVert^2,
	\end{IEEEeqnarray}
	while Encoder 1 communicates at a normalized rate of
	\begin{IEEEeqnarray}{c}
		r_1(\bm{h}_{ri},h_d)=\frac{2}{\ln(2)}\rb{\Big\lvert\bm{h}_{ri}^\intercal e^{j\tilde{\pmb{\theta}}}+h_d\Big\rvert^2-\lVert \bm{h}_{ri}\rVert^2}.
	\end{IEEEeqnarray}
\end{corollary}
\begin{IEEEproof}
	See Appendix \ref{app:proof_low_power_corner_points}.
\end{IEEEproof}

The asymmetry between Encoder 1 and Encoder 2 revealed by \cref{cor:low_power_corner_points} stems from the fact the, in order for Encoder 1 to obtain its maximum rate in the low-power regime, Encoder 2 needs to steer its phases according to the beamforming solution \eqref{eq:beamforming}. This in turn makes it impossible to encode additional information for Encoder 2. In contrast, Encoder 2's maximum rate can be obtained as long as Encoder 1's signal is transmitted at the maximum power and can be decoded while treating the modulation of the \ac{RIS}'s phases by Encoder 2 as a nuisance.

We finally remark that \cref{cor:high_snr} and \cref{cor:low_power_corner_points} imply that time-sharing, which would yield a triangular rate region, is suboptimal in both high- and low-power regimes.
This is in contrast to the multiplicative \ac{MAC} studied in \cite{pillai2011on} that assumes two standard active encoders with separate power constraints.

\section{Examples}
In this section, we provide numerical examples for the capacity region derived in \cref{sec:capacity_region}. For the phase response set, we consider $A$ uniformly spaced phases in the set $\mathcal A=\{0,2\pi/A,\ldots,2\pi(A-1)/A\}$, whereas, for the input constellation, we consider \ac{ASK} and \ac{PSK} modulations. In addition, we assume a channel vector $\bm{h}_d$ with elements having amplitude $1$, and a channel matrix $\bm{H}_{ri}$ with elements having amplitude $\alpha>0$, where $\alpha$ denotes the path-loss ratio between the reflected and direct paths. The phases of $\bm{H}_{ri}$ and $\bm{h}_d$ used in this section are summarized in \cref{table_example}.
\begin{table}[!t]
	\renewcommand{\arraystretch}{1.3}
	\caption{Phases of $\bm{H}_{ri}$ and $\bm{h}_d$ used for the numerical examples}
	\label{table_example}
	\centering
	\begin{tabular}{|c|c|c|}
		\hline
		\bfseries Figure  & \bfseries $\angle \bm{H}_{ri}$ [rad] & \bfseries $\angle\bm{h}_{d}$ [rad]\\
		\hline
		\ref{fig:cap_low_snr} & $\begin{pmatrix}
			1.11 & 0.71 & 2.92 & -2.29\\
			2.52 & -0.72& 2.21 & 2.1
		\end{pmatrix}$
		& $\begin{pmatrix}
			3.11 \\
			1.39
		\end{pmatrix}$
		\\
		\hline
		\ref{fig:cap_high_snr} & $\begin{pmatrix}
			-2.63 & -1.22 & -2.92 & -1.52\\
			1.85 & 0.36& -0.87 & -2.59
		\end{pmatrix}$
		& $\begin{pmatrix}
			2.82 \\
			2.32
		\end{pmatrix}$
		\\
		\hline
	\end{tabular}
\end{table}
Furthermore, the expectation over Gaussian random vectors, e.g., $\mathbf{z}$ in \cref{prop:capacity_region}, is evaluated via a Monte Carlo empirical averages. 

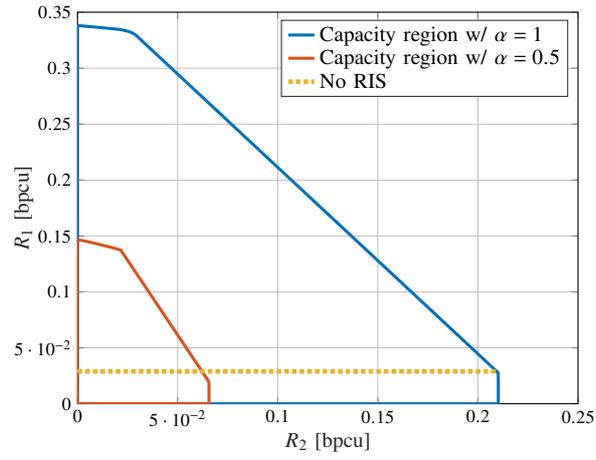
\begin{figure}[!t]
	\centering
	\resizebox {0.5\columnwidth} {!} {	
		\definecolor{mycolor1}{rgb}{0.00000,0.44700,0.74100}%
\definecolor{mycolor2}{rgb}{0.85000,0.32500,0.09800}%
\definecolor{mycolor3}{rgb}{0.92900,0.69400,0.12500}%

\begin{tikzpicture}

\begin{axis}[%
width=4.521in,
height=3.566in,
at={(0.758in,0.481in)},
scale only axis,
xmin=0,
xmax=0.25,
xlabel style={font=\color{white!15!black}},
xlabel={{\large $R_2$ [bpcu]}},
ymin=0,
ymax=0.35,
ylabel style={font=\color{white!15!black}},
ylabel={{\large $R_1$ [bpcu]}},
axis background/.style={fill=white},
xmajorgrids,
ymajorgrids,
xtick={0,0.05,0.1,0.15,0.2,0.25},
legend style={legend cell align=left, align=left, draw=white!15!black}
]
\addplot [color=mycolor1, line width=2.0pt]
  table[row sep=crcr]{%
0	0\\
0.210176878810206	0\\
0.210176878810206	0.00068243062870188\\
0.2101768534749	0.0217286061731228\\
0.210176838531623	0.0217295740494734\\
0.21002616844009	0.027761476051503\\
0.210026008777404	0.0277658177489122\\
0.210025796928842	0.0277702645858695\\
0.21002552406407	0.0277749026433822\\
0.21002519747128	0.0277795528536027\\
0.210024794871258	0.027784467712562\\
0.210024328139401	0.0277894350671617\\
0.210023782934173	0.0277945694997088\\
0.210023162483854	0.027799796393098\\
0.210022452055652	0.0278051996767119\\
0.210021661951314	0.0278106677648942\\
0.210020753275218	0.0278164239477916\\
0.210019749266663	0.0278222739530194\\
0.210018628165368	0.0278283126058012\\
0.21001738274156	0.0278345379359783\\
0.210016010134963	0.0278409282683523\\
0.210014488404704	0.0278475469200812\\
0.210012829867154	0.0278543059884884\\
0.210010992718058	0.0278613384107094\\
0.210009005952307	0.0278684979918751\\
0.210006801277995	0.0278759945333253\\
0.210004414953621	0.0278836612379743\\
0.029363660021763	0.329126962112436\\
0.0292890544376907	0.3292492864463\\
0.0292114901341285	0.329371142268633\\
0.0291288894189461	0.329495549850975\\
0.0290418174009428	0.329621299218795\\
0.0289501485036818	0.329748293339785\\
0.0288527464708133	0.329877752279009\\
0.0287505143664366	0.330008139414844\\
0.0286434155077444	0.330139267751327\\
0.0285305165607053	0.330272001354941\\
0.0284108763588651	0.33040706626245\\
0.0282849582324363	0.330543562394179\\
0.0281528439814513	0.330681096256986\\
0.0280137984171551	0.330820118496435\\
0.0278675015277843	0.33096060179558\\
0.0277134961102488	0.331102626042206\\
0.0275517706878312	0.331245849512561\\
0.027381848234056	0.331390348736366\\
0.0272027343761647	0.331536578880106\\
0.0270152564950972	0.331683501051073\\
0.0268183730494576	0.331831586788547\\
0.0266109585591958	0.331981270264281\\
0.026394297474436	0.332131251264101\\
0.0261661259700987	0.332282713425577\\
0.0259274541683747	0.332434580375661\\
0.0256773182103363	0.332587093471207\\
0.0254142834814046	0.332740690922155\\
0.0251399342111855	0.332894049054894\\
0.024852102322297	0.333047988479217\\
0.0245512382967759	0.33320184350117\\
0.0242371888375308	0.33335530780819\\
0.0239084116714983	0.333508720735346\\
0.0235655427004602	0.33366136397096\\
0.0232083576337381	0.333812958070825\\
0.0228356790561106	0.333963601927774\\
0.0224485965235419	0.334112473058396\\
0.0220447266785295	0.33426008927428\\
0.0216262969993255	0.334405252748179\\
0.0211921636872776	0.334548019753538\\
0.0207496500306767	0.334685816886571\\
0.0207496498706656	0.334685816915874\\
0.0207496473851454	0.334685817351152\\
0.01007747349776	0.336509465485145\\
0.000682430628702768	0.337999935604814\\
0	0.337999935604814\\
0	0\\
};
\addlegendentry{{\large Capacity region w/ $\alpha=1$}}

\addplot [color=mycolor2, line width=2.0pt]
  table[row sep=crcr]{%
0	0\\
0.0656753659189695	0\\
0.0656753659189695	0.000682430628702768\\
0.0656753146034235	0.0178942710396881\\
0.0656752133082659	0.0179008310396758\\
0.0656750383676998	0.0179075572088241\\
0.0656747902817085	0.0179143091713692\\
0.0656744600946122	0.0179212378131113\\
0.0656740449357174	0.0179282949227386\\
0.0655622901219934	0.0194371994760032\\
0.0655320983307761	0.019829666517774\\
0.0655309378448758	0.0198420146216538\\
0.0655295965018574	0.0198546449371371\\
0.0655280999314072	0.0198672646310185\\
0.0655263424261454	0.0198806438767836\\
0.0655243684678748	0.0198942900232764\\
0.0655222353116494	0.0199077888519317\\
0.0655198161836359	0.0199218832320525\\
0.0655171575482152	0.0199362019289993\\
0.0217039459320505	0.137128269303994\\
0.0216946677004706	0.13715239658462\\
0.0216855292576499	0.137175005032251\\
0.0216763697380413	0.13719658327715\\
0.0216672970645013	0.137216970381495\\
0.0216582931359071	0.137236277015267\\
0.0216491711460778	0.137254950703662\\
0.0216401088638452	0.137272671045288\\
0.0216311570665186	0.137289402433008\\
0.0216223063402947	0.13730523273978\\
0.0216134713071057	0.137320358126901\\
0.0216047118796383	0.137334721855043\\
0.0215961298471017	0.137348208101239\\
0.0215873426215074	0.137361448639963\\
0.0215787847729492	0.137373807629419\\
0.0215703359863881	0.137385513872512\\
0.0215617702604765	0.137396898246412\\
0.0215535476806501	0.137407390377137\\
0.0215451143186338	0.137417720283465\\
0.0215369353874713	0.137427333278489\\
0.0215286733613316	0.137436661145612\\
0.0215205169321666	0.137445501652584\\
0.0215127185052513	0.137453621272277\\
0.0215046507353143	0.137461691129301\\
0.0214968707450023	0.137469165976364\\
0.0214889205915241	0.137476502309414\\
0.0214812059802854	0.137483335294688\\
0.0214735819298086	0.137489820598207\\
0.0214659594101523	0.137496046247969\\
0.0214584300210454	0.137501949121551\\
0.0214508200398842	0.137507672315228\\
0.0214434015795062	0.137513022883072\\
0.0214361381304262	0.137518048179572\\
0.0214291168080507	0.137522709837888\\
0.0214214823909771	0.137527565461006\\
0.0214145488117317	0.137531787171246\\
0.0214075095311537	0.137535894052702\\
0.0214007789927413	0.137539655636467\\
0.0213937819587753	0.137543398667442\\
0.0213870298454837	0.137546852048299\\
0.0180245087094151	0.139189022143416\\
0.0143438012827204	0.140906328726258\\
0.0104151440179518	0.142655254936173\\
0.00624320219648045	0.144425789271174\\
0.00421774913511941	0.145254293187366\\
0.0038473492481601	0.145396681697856\\
0.00342734212043583	0.145550122456772\\
0.00295328890871138	0.145714505303013\\
0.00241280014959422	0.145892165823681\\
0.00179667673687556	0.14608385605467\\
0.00108438787337128	0.146293236386569\\
0.000682430628702324	0.146406347564662\\
0	0.146406347564662\\
0	0\\
};
\addlegendentry{{\large Capacity region w/ $\alpha=0.5$}}

\addplot [color=mycolor3, dotted, line width=3.0pt]
  table[row sep=crcr]{%
0	0.0289228730504312\\
0.210176878810206	0.0289228730504312\\
};
\addlegendentry{{\large No RIS}}
\end{axis}
\end{tikzpicture}%
	}
	\caption{Capacity region for $P=-20$ dB, $N=2$, $K=4$, $A=2$, $m=2$, and BPSK input constellation. The dashed line illustrates the capacity of Encoder 1 for a channel with no \ac{RIS}.}
	\label{fig:cap_low_snr}
\end{figure}

In \cref{fig:cap_low_snr}, we plot the capacity region for an average power constraint of $P=-20$ dB, $N=2$ receiver antennas, $K=4$ RIS elements, $A=2$ available phase shifts, a symbol-to-RIS control rate $m=2$, input constellation given by BPSK, i.e., $\mathcal S=\{-1,1\}$, and a path-loss ratio of $\alpha=0.5$ or $\alpha=1$.
In addition, we plot for reference the maximum rate achievable by Encoder 1 for a channel with no \ac{RIS}, i.e., for $\bm{H}_{ri}=\mathbf{0}$.
By comparing with the capacity of the channel with no \ac{RIS}, \cref{fig:cap_low_snr} illustrates the two roles of the \ac{RIS}: The \ac{RIS} can be used to increase the rate of Encoder 1 by beamforming the transmitted signal, and it can enable communication from a passive secondary user. 
In this regard, \cref{fig:cap_low_snr} demonstrates that the insights obtained in \cref{cor:low_power_corner_points} by studying the low-power regime carry over to more general conditions. In particular, the maximum rate for Encoder 1 is achieved if and only if Encoder 2 does not communicate, while Encoder 2's maximum rate can coexist with a non-zero rate for Encoder 1.

\begin{figure}[!t]
	\centering
	\resizebox {0.5\columnwidth} {!} {	
		\definecolor{mycolor1}{rgb}{0.00000,0.44700,0.74100}%
\definecolor{mycolor2}{rgb}{0.92900,0.69400,0.12500}%
\begin{tikzpicture}

\begin{axis}[%
width=4.521in,
height=3.566in,
at={(0.758in,0.481in)},
scale only axis,
xmin=0,
xmax=4.5,
xlabel style={font=\color{white!15!black}},
xlabel={{\large $R_2$ [bpcu]}},
ymin=0,
ymax=2.5,
ylabel style={font=\color{white!15!black}},
ylabel={{\large $R_1$ [bpcu]}},
axis background/.style={fill=white},
xmajorgrids,
ymajorgrids,
legend style={legend cell align=left, align=left, draw=white!15!black}
]
\addplot [color=mycolor1, line width=2.0pt]
  table[row sep=crcr]{%
0	0\\
4.00061238745425	0\\
4.00061238745425	2.00061238745427\\
0	2.00061238745427\\
0	0\\
};
\addlegendentry{{\large Capacity region}}

\addplot [color=mycolor2, dotted, line width=3.0pt]
  table[row sep=crcr]{%
0	2.00061238745427\\
4.00061238745425	2.00061238745427\\
};
\addlegendentry{{\large No \ac{RIS}}}

\end{axis}
\end{tikzpicture}%
	}
	\caption{Capacity region for $P=40$ dB, $N=2$, $K=4$, $A=2$, $m=1$, and 4-ASK input constellation. The dashed line illustrates the capacity of Encoder 1 for a channel with no \ac{RIS}.}
	\label{fig:cap_high_snr}
\end{figure}
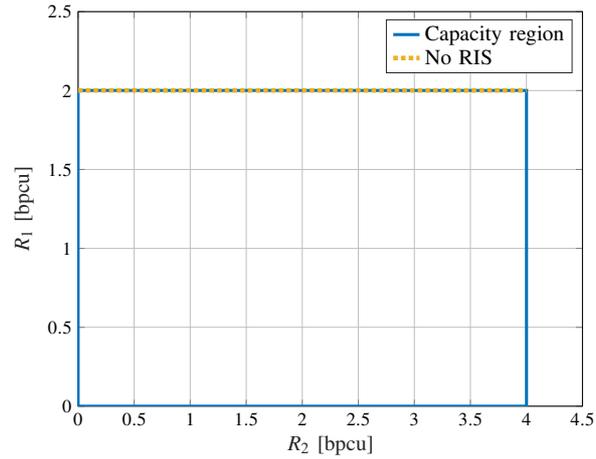

In contrast, by \cref{cor:high_snr}, for sufficiently high power $P$, both encoders can communicate with the decoder at their respective maximum rates. This is verified by \cref{fig:cap_high_snr}, where we plot the capacity region for an average power constraint of $P=40$ dB, $N=2$ receiver antennas, $K=4$ RIS elements, $A=2$ available phase shifts, a symbol-to-RIS control rate $m=1$, input constellation given by 4-\ac{ASK}, i.e., $\mathcal S=\{\sigma,3\sigma,5\sigma,7\sigma\}$ with $\sigma=1/\sqrt{21}$, and a path-loss ratio of $\alpha=1$.
Although Encoder 1 does not gain from the existence of the \ac{RIS} in the high-power regime, the \ac{RIS} enables \ac{MU} communication with a single transmitter in a manner that resembles the single-\ac{RF} \ac{MIMO} system \cite{li2020single,karasik2020adaptive}.

\section{Conclusion}
In this work, we have studied the finite-input capacity region of an \ac{RIS}-aided \ac{MU} communication system, in which the \ac{RIS} is not used solely for increasing the rate of an active encoder, but also for enabling communication for a secondary passive encoder. The fundamental trade-offs between the rates of the two encoders were characterized. It was shown that, for sufficiently high power, both users can communicate at their respective maximum rates. Furthermore, in the low-power regime, the maximum rate for the active encoder is achieved if and only if the passive encoder does not communicate, while the passive encoder's maximum rate can coexist with a non-zero rate for the active encoder. Finally, time-sharing was demonstrated to be suboptimal.

\appendix
\subsection{Proof of Proposition \ref{prop:capacity_region}}\label{app:proof_capacity_region}
The model \eqref{eq:vec_channel} can be viewed as a \ac{MAC} with inputs $(\mathbf{s},\pmb{\uptheta})$ and output $\mathbf{y}$. Therefore, it follows from the capacity region of the \ac{MAC} \cite[Thm. 4.2]{el2011network} that $\mathcal{C}(\bm{H}_{ri},\bm{h}_d)$ is the convex hull of the union of regions $\tilde{\mathcal{R}}(p_{\mathbf{s}},p_{\pmb{\uptheta}},\bm{H}_{ri},\bm{h}_d)$ over all input distributions $p_{\mathbf{s}}(\bm{s})$ and $p_{\pmb{\uptheta}}(\pmb{\theta})$ such that $\mathbb{E}[\mathbf{s}^\ast\mathbf{s}]\leq m$, where $\tilde{\mathcal{R}}(p_{\mathbf{s}},p_{\pmb{\uptheta}},\bm{H}_{ri},\bm{h}_d)$ is the set of rate pairs $\rb{R_1(\bm{H}_{ri},\bm{h}_d),R_2(\bm{H}_{ri},\bm{h}_d)}$ such that inequalities
\begin{IEEEeqnarray}{rCl}\label{eq:app_capacity_region}
	R_1(\bm{H}_{ri},\bm{h}_d)&\leq& \frac{1}{m}I(\mathbf{s};\mathbf{y}|\pmb{\uptheta}),\IEEEyesnumber\IEEEyessubnumber\\
	R_2(\bm{H}_{ri},\bm{h}_d)&\leq& \frac{1}{m}I(\pmb{\uptheta};\mathbf{y}|\mathbf{s}),\IEEEyessubnumber\\
	\text{and }R_1(\bm{H}_{ri},\bm{h}_d)+R_2(\bm{H}_{ri},\bm{h}_d)&\leq&\frac{1}{m}I(\mathbf{s},\pmb{\uptheta};\mathbf{y})\IEEEyessubnumber
\end{IEEEeqnarray}
hold. Since inputs $\mathbf{s}$ and $\pmb{\uptheta}$ are selected from finite sets, the mutual information $I(\mathbf{s};\mathbf{y}|\pmb{\uptheta})$ in \eqref{eq:app_capacity_region} can be written as (see, e.g., \cite[App. A]{karasik2020adaptive})
\begin{IEEEeqnarray}{rCl}\label{eq:app_mi1}
	I(\mathbf{s};\mathbf{y}|\pmb{\uptheta})&=&-NM\log_2(e)
	-\int_{\mathbb{C}^{Nm\times 1}}\nspace{mq}p_{\mathbf{z}}(\bm{z})\sum_{\bm{s}_1\in\mathcal{S}^{m\times 1}}\nspace{!}\nspace{!}p_{\mathbf{s}}(\bm{s}_1)\sum_{\pmb{\theta}_1\in\mathcal{A}^{K\times 1}}\nspace{!}\nspace{!}p_{\pmb{\uptheta}}(\pmb{\theta}_1)
	\log_2\rb{\sum_{\bm{s}_2\in\mathcal{S}^{m\times 1}}\nspace{!}\nspace{!}p_{\mathbf{s}}(\bm{s}_2)e^{u_1}}\diff{\bm{z}}\IEEEeqnarraynumspace
\end{IEEEeqnarray}
with $\mathbf{z}\sim\mathcal{CN}(\bm{0},\bm{I}_{Nm})$ and where we have defined the scalar
\begin{IEEEeqnarray}{c}
	u_1\triangleq -\Big\lVert\bm{z}+\sqrt{P}\rb{\bm{H}_{ri}e^{j\pmb{\theta}_1}+\bm{h}_d}^{m\otimes}\rb{\bm{s}_1-\bm{s}_2}\Big\rVert^2.
\end{IEEEeqnarray}
By applying the conditional \ac{CGF} definition in \eqref{eq:def_cond_cgf_rand} to \eqref{eq:app_mi1}, we get
\begin{IEEEeqnarray}{c}
	I(\mathbf{s};\mathbf{y}|\pmb{\uptheta})=-Nm\log_2(e)-\kappa(\mathrm{u}_1|\mathbf{s}_1,\pmb{\uptheta}_1,\mathbf{z}).
\end{IEEEeqnarray}
Similarly, we also have
\begin{IEEEeqnarray}{rCl}
	I(\pmb{\uptheta};\mathbf{y}|\mathbf{s})&=& -Nm\log_2(e)-\kappa(\mathrm{u}_2|\mathbf{s}_1,\pmb{\uptheta}_1,\mathbf{z}),\IEEEyesnumber\IEEEyessubnumber\\
	I(\mathbf{s},\pmb{\uptheta};\mathbf{y})&=&-Nm\log_2(e)-\kappa(\mathrm{u}_3|\mathbf{s}_1,\pmb{\uptheta}_1,\mathbf{z})\IEEEyessubnumber.
\end{IEEEeqnarray}
Therefore, the region $\tilde{\mathcal{R}}(p_{\mathbf{s}},p_{\pmb{\uptheta}},\bm{H}_{ri},\bm{h}_d)$ in \eqref{eq:app_capacity_region} is identical to the region $\mathcal{R}(p_{\mathbf{s}},p_{\pmb{\uptheta}},\bm{H}_{ri},\bm{h}_d)$ in \eqref{eq:capacity_region}.

\subsection{Proof of Corollary \ref{cor:high_snr}}\label{app:proof_high_snr}
The inclusion $\mathcal{C}(\bm{H}_{ri},\bm{h}_d)\subseteq \overline{\mathcal C}$ is trivial since, for all input distributions $p_{\mathbf{s}}(\bm{s})$ and $p_{\pmb{\uptheta}}(\pmb{\theta})$ with $\bm{s}\in\mathcal S^{m\times 1}$ and $\pmb{\theta}\in\mathcal A^{K\times 1}$ we have $H(\mathbf{s})\leq m\log_2(S)$ and $H(\pmb{\uptheta})\leq K\log_2(A)$. In addition, in the high-power regime, we have the limits
\begin{IEEEeqnarray}{rCl}\label{eq:app_high_snr_limit1}
	I(\mathbf{s};\mathbf{y}|\pmb{\uptheta})&\xrightarrow[P\rightarrow\infty]{}&H(\mathbf{s})\leq m\log_2(S),\IEEEyesnumber\IEEEyessubnumber\\
	I(\pmb{\uptheta};\mathbf{y}|\mathbf{s})&\xrightarrow[P\rightarrow\infty]{}&H(\pmb{\uptheta})\leq K\log_2(A),\IEEEyessubnumber
\end{IEEEeqnarray}
where equality is achieved for a uniform distributions $p_{\mathbf{s}}(\bm{s})$ and $p_{\pmb{\uptheta}}(\pmb{\theta})$.
Next, note that the noiseless received signal $\mathbf{y}(t)-\mathbf{z}(t)$ in \eqref{eq:vec_channel} takes values from a discrete set. Furthermore, since channel matrix $\mathbf{H}_{ri}$ and channel vector $\mathbf{h}_d$ are drawn from a continuous distribution, almost surely, for all $t\in[n/m]$, there exist unique inputs $\hat{\bm{s}}(t)\in\mathcal{S}^{m\times 1}$ and $\hat{\pmb{\theta}}(t)\in\mathcal{A}^{K\times 1}$ such that (see, e.g., \cite{motahari2014real})
\begin{IEEEeqnarray}{c}
	\mathbf{y}(t)-\mathbf{z}(t)=\sqrt{P}\rb{\bm{H}_{ri}e^{j\hat{\pmb{\theta}}(t)}+\bm{h}_d}^{m\otimes}\hat{\bm{s}}(t).
\end{IEEEeqnarray}
Therefore, for all input distributions $p_{\mathbf{s}}(\bm{s})$ and $p_{\pmb{\uptheta}}(\pmb{\theta})$, the transmitted signal $\mathbf{s}(t)$ and reflection pattern $\pmb{\uptheta}(t)$ can be correctly jointly decoded in the high-power regime, i.e., we have the limit
\begin{IEEEeqnarray}{c}\label{eq:app_high_snr_limit2}
	I(\mathbf{s},\pmb{\uptheta};\mathbf{y})\xrightarrow[P\rightarrow\infty]{}H(\mathbf{s})+H(\pmb{\uptheta})\leq m\log_2(S)+K\log_2(A).
\end{IEEEeqnarray}
Let $\rb{R_1^u(\bm{H}_{ri},\bm{h}_d),R_2^u(\bm{H}_{ri},\bm{h}_d)}\in \mathcal{C}(\bm{H}_{ri},\bm{h}_d)$ be the rate pair achieved using uniform distributions $p_{\mathbf{s}}(\bm{s})$ and $p_{\pmb{\uptheta}}(\pmb{\theta})$.
It hence follows from the region in \eqref{eq:app_capacity_region} and limits \eqref{eq:app_high_snr_limit1} and \eqref{eq:app_high_snr_limit2} that, almost surely, we have the limits
\begin{IEEEeqnarray}{rCl}
	\lim_{P\rightarrow\infty}R_1^u(\bm{H}_{ri},\bm{h}_d)&=&\log_2(S),\IEEEyesnumber\IEEEyessubnumber\\
	\lim_{P\rightarrow\infty}R_2^u(\bm{H}_{ri},\bm{h}_d)&=&\frac{K}{m}\log_2(A)\IEEEyessubnumber.
\end{IEEEeqnarray}

\subsection{Proof of Proposition \ref{prop:low_snr}}\label{app:proof_low_snr}
For input distributions $p_{\mathrm{s}}(s)$ and $p_{\pmb{\uptheta}}(\pmb{\theta})$, let functions $\tilde{R}_\ell(P,\bm{h}_{ri},h_d)$, $\ell\in\{1,2,3\}$, be defined as
\begin{IEEEeqnarray}{c}
	\tilde{R}_\ell(P,\bm{h}_{ri},h_d)\triangleq-\log_2(e)-\kappa(\mathrm{u}_\ell|\mathrm{s}_1,\pmb{\uptheta}_1,\mathrm{z}),
\end{IEEEeqnarray}
where $\kappa(\mathrm{u}_\ell|\mathrm{s}_1,\pmb{\uptheta}_1,\mathrm{z})$ are the conditional \ac{CGF}s in \cref{prop:capacity_region} for the special case in which $N=m=1$. By calculating the derivative of $\tilde{R}_\ell(P,\bm{h}_{ri},h_d)$ with respect to the power $P$ and taking the limit $P\rightarrow0$, we get
\begin{IEEEeqnarray}{c}
	\lim_{P\rightarrow 0}\frac{\partial \tilde{R}_\ell(P,\bm{h}_{ri},h_d)}{\partial P}=\frac{\mathbb E[\underline{\mathrm{u}}_\ell]}{\ln(2)},
\end{IEEEeqnarray}
where random variables $\underline{\mathrm{u}}_\ell$ are defined in \eqref{eq:def_u_low_power}. Therefore, it follows from \cref{prop:capacity_region} that the normalized rate pairs $\rb{r_1(\bm{h}_{ri},h_d),r_2(\bm{h}_{ri},h_d)}$ satisfy
\begin{IEEEeqnarray}{rCl}
	r_\ell(\bm{h}_{ri},h_d)&=&\lim_{P\rightarrow 0}\frac{R_\ell(\bm{h}_{ri},h_d)}{P} \IEEEnonumber\\
	&\leq&\lim_{P\rightarrow 0}\frac{\tilde{R}_\ell(P,\bm{h}_{ri},h_d)}{P} \IEEEnonumber\\
	&=&\lim_{P\rightarrow 0}\frac{\partial \tilde{R}_\ell(P,\bm{h}_{ri},h_d)}{\partial P}\IEEEnonumber\\
	&=&\frac{\mathbb E[\underline{\mathrm{u}}_\ell]}{\ln(2)},\quad\ell\in\{1,2\},
\end{IEEEeqnarray}
and similarly we have
\begin{IEEEeqnarray}{c}
	r_1(\bm{h}_{ri},h_d)+r_2(\bm{h}_{ri},h_d)\leq\frac{\mathbb E[\underline{\mathrm{u}}_3]}{\ln(2)}.
\end{IEEEeqnarray}

\subsection{Proof of Corollary \ref{cor:low_power_corner_points}}\label{app:proof_low_power_corner_points}
Since $\mathrm{s}_1$, $\mathrm{s}_2$, and $\pmb{\uptheta}_1$ in \cref{prop:low_snr} are all independent, we have
\begin{IEEEeqnarray}{rCl}
	\mathbb E[\underline{\mathrm{u}}_1]&=&\mathbb E\sqb{\Big\lvert\rb{\bm{h}_{ri}^\intercal e^{j\pmb{\uptheta}_1}+h_d}\rb{\mathrm{s}_1-\mathrm{s}_2}\Big\rvert^2}\label{eq:app_low_power_bound1}\IEEEnonumber\\
	&=&\mathbb E\sqb{\Big\lvert\bm{h}_{ri}^\intercal e^{j\pmb{\uptheta}_1}+h_d\Big\rvert^2}\mathbb E\sqb{\lvert\mathrm{s}_1-\mathrm{s}_2\rvert^2}\IEEEnonumber\\
	&\leq& 2\Big\lvert\bm{h}_{ri}^\intercal e^{j\tilde{{\uptheta}}}+h_d\Big\rvert^2.
\end{IEEEeqnarray}
Similarly, we have the upper bounds
\begin{IEEEeqnarray}{rCl}
	\mathbb E[\underline{\mathrm{u}}_2]&\leq&2\lVert \bm{h}_{ri}\rVert^2,\IEEEyesnumber\IEEEyessubnumber\label{eq:app_low_power_bound2}\\
	\mathbb E[\underline{\mathrm{u}}_3]&\leq&2\Big\lvert\bm{h}_{ri}^\intercal e^{j\tilde{\pmb{\uptheta}}}+h_d\Big\rvert^2.\IEEEyessubnumber\label{eq:app_low_power_bound3}
\end{IEEEeqnarray}
Equality in \eqref{eq:app_low_power_bound1} and \eqref{eq:app_low_power_bound3} is achieved for fixed \ac{RIS} reflection pattern $\pmb{\uptheta}=\tilde{\pmb{\theta}}$ with probability one and uniform input distribution $p_{\mathrm{s}}(s)=1/S$. Furthermore, since the upper bounds in \eqref{eq:app_low_power_bound1} and \eqref{eq:app_low_power_bound3} are equal, Encoder 1 can achieve the maximum normalized rate if and only if $\pmb{\uptheta}=\tilde{\pmb{\theta}}$ with probability one.
In contrast, equality in \eqref{eq:app_low_power_bound2} is achieved for uniform phase-shift distribution $p_{\pmb{\uptheta}}(\pmb{\theta})=1/A^K$ and any input distribution $p_{\mathrm{s}}(s)$ for which $\mathbb E[|\mathrm{s}|^2]=1$. 
That is, Encoder 2 can achieve the maximum normalized rate, while Encoder 1 transmits at a positive normalized rate. 

\bibliographystyle{IEEEtran}
\bibliography{IEEEabrv,myBib}

\begin{thebibliography}{10}
\providecommand{\url}[1]{#1}
\csname url@samestyle\endcsname
\providecommand{\newblock}{\relax}
\providecommand{\bibinfo}[2]{#2}
\providecommand{\BIBentrySTDinterwordspacing}{\spaceskip=0pt\relax}
\providecommand{\BIBentryALTinterwordstretchfactor}{4}
\providecommand{\BIBentryALTinterwordspacing}{\spaceskip=\fontdimen2\font plus
\BIBentryALTinterwordstretchfactor\fontdimen3\font minus
  \fontdimen4\font\relax}
\providecommand{\BIBforeignlanguage}[2]{{%
\expandafter\ifx\csname l@#1\endcsname\relax
\typeout{** WARNING: IEEEtran.bst: No hyphenation pattern has been}%
\typeout{** loaded for the language `#1'. Using the pattern for}%
\typeout{** the default language instead.}%
\else
\language=\csname l@#1\endcsname
\fi
#2}}
\providecommand{\BIBdecl}{\relax}
\BIBdecl

\bibitem{renzo2020Smart}
M.~{Di Renzo}, A.~{Zappone}, M.~{Debbah}, M.~{Alouini}, C.~{Yuen}, J.~D.
  {Rosny}, and S.~{Tretyakov}, ``Smart radio environments empowered by
  reconfigurable intelligent surfaces: How it works, state of research, and
  road ahead,'' \emph{{IEEE} J. Sel. Areas Commun.}, 2020.

\bibitem{renzo202Reconfigurable}
M.~{Di Renzo}, K.~{Ntontin}, J.~{Song}, F.~H. {Danufane}, X.~{Qian},
  F.~{Lazarakis}, J.~{De Rosny}, D.~{Phan-Huy}, O.~{Simeone}, R.~{Zhang},
  M.~{Debbah}, G.~{Lerosey}, M.~{Fink}, S.~{Tretyakov}, and S.~{Shamai},
  ``Reconfigurable intelligent surfaces vs. relaying: Differences,
  similarities, and performance comparison,'' \emph{IEEE Open Journal of the
  Communications Society}, vol.~1, pp. 798--807, 2020.

\bibitem{wu2020towards}
Q.~{Wu} and R.~{Zhang}, ``Towards smart and reconfigurable environment:
  Intelligent reflecting surface aided wireless network,'' \emph{{IEEE} Commun.
  Mag.}, vol.~58, no.~1, pp. 106--112, 2020.

\bibitem{liu2020reconfigurable}
Y.~Liu, X.~Liu, X.~Mu, T.~Hou, J.~Xu, Z.~Qin, M.~Di~Renzo, and N.~Al-Dhahir,
  ``Reconfigurable intelligent surfaces: Principles and opportunities,''
  \emph{arXiv preprint arXiv:2007.03435}, 2020.

\bibitem{renzo2019smart}
M.~Di~Renzo, M.~Debbah, D.-T. Phan-Huy, A.~Zappone, M.-S. Alouini, C.~Yuen,
  V.~Sciancalepore, G.~C. Alexandropoulos, J.~Hoydis, H.~Gacanin, J.~de~Rosny,
  A.~Bounceu, G.~Lerosey, and M.~Fink, ``Smart radio environments empowered by
  {AI} reconfigurable meta-surfaces: An idea whose time has come,''
  \emph{EURASIP J. Wireless Commun. Netw.}, pp. 1--20, 2019.

\bibitem{yuan2020reconfigurable}
X.~Yuan, Y.-J. Zhang, Y.~Shi, W.~Yan, and H.~Liu,
  ``Reconfigurable-intelligent-surface empowered {6G} wireless communications:
  Challenges and opportunities,'' \emph{arXiv preprint arXiv:2001.00364}, 2020.

\bibitem{wu2020intelligent}
Q.~Wu, S.~Zhang, B.~Zheng, C.~You, and R.~Zhang, ``Intelligent reflecting
  surface aided wireless communications: A tutorial,'' \emph{arXiv preprint
  arXiv:2007.02759}, 2020.

\bibitem{guo2019weighted2}
H.~{Guo}, Y.~C. {Liang}, J.~{Chen}, and E.~G. {Larsson}, ``Weighted sum-rate
  maximization for intelligent reflecting surface enhanced wireless networks,''
  in \emph{Proc. IEEE Global Conf. Communications (GLOBECOM)}, 2019, pp. 1--6.

\bibitem{zhang2020reconfigurable}
H.~Zhang, B.~Di, Z.~Han, H.~V. Poor, and L.~Song, ``Reconfigurable intelligent
  surface assisted multi-user communications: How many reflective elements do
  we need?'' \emph{arXiv preprint arXiv:2012.10736}, 2020.

\bibitem{abrardo2020intelligent}
A.~Abrardo, D.~Dardari, and M.~Di~Renzo, ``Intelligent reflecting surfaces:
  Sum-rate optimization based on statistical {CSI},'' \emph{arXiv preprint
  arXiv:2012.10679}, 2020.

\bibitem{mu2020joint}
X.~Mu, Y.~Liu, L.~Guo, J.~Lin, and R.~Schober, ``Joint deployment and multiple
  access design for intelligent reflecting surface assisted networks,''
  \emph{arXiv preprint arXiv:2005.11544}, 2020.

\bibitem{huang2019reconfigurable}
C.~{Huang}, A.~{Zappone}, G.~C. {Alexandropoulos}, M.~{Debbah}, and C.~{Yuen},
  ``Reconfigurable intelligent surfaces for energy efficiency in wireless
  communication,'' \emph{{IEEE} Trans. Wireless Commun.}, vol.~18, no.~8, pp.
  4157--4170, 2019.

\bibitem{han2020intelligent}
H.~{Han}, J.~{Zhao}, D.~{Niyato}, M.~D. {Renzo}, and Q.~{Pham}, ``Intelligent
  reflecting surface aided network: Power control for physical-layer
  broadcasting,'' in \emph{Proc. IEEE Int. Conf. Communications (ICC)}, 2020,
  pp. 1--7.

\bibitem{zhang2020physical}
J.~Zhang, H.~Du, Q.~Sun, D.~W.~K. Ng, and B.~Ai, ``Physical layer security
  enhancement with reconfigurable intelligent surface-aided networks,''
  \emph{arXiv preprint arXiv:2012.00269}, 2020.

\bibitem{zhou2020fairness}
G.~Zhou, C.~Pan, H.~Ren, K.~Wang, and M.~Di~Renzo, ``Fairness-oriented multiple
  {RIS}s-aided {MmWave} transmission: Stochastic optimization approaches,''
  \emph{arXiv preprint arXiv:2012.06103}, 2020.

\bibitem{yang2020novel}
L.~Yang, F.~Meng, M.~O. Hasna, and E.~Basar, ``A novel {RIS}-assisted
  modulation scheme,'' \emph{arXiv preprint arXiv:2011.09019}, 2020.

\bibitem{pillai2011on}
S.~R.~B. {Pillai}, ``On the capacity of multiplicative multiple access channels
  with awgn,'' in \emph{Proc. IEEE Inform. Theory Workshop (ITW)}, 2011, pp.
  452--456.

\bibitem{liu2018backscatter}
W.~{Liu}, Y.~{Liang}, Y.~{Li}, and B.~{Vucetic}, ``Backscatter multiplicative
  multiple-access systems: Fundamental limits and practical design,''
  \emph{{IEEE} Trans. Wireless Commun.}, vol.~17, no.~9, pp. 5713--5728, 2018.

\bibitem{karasik2020adaptive}
R.~Karasik, O.~Simeone, M.~Di~Renzo, and S.~Shamai, ``Adaptive coding and
  channel shaping through reconfigurable intelligent surfaces: An
  information-theoretic analysis,'' \emph{arXiv preprint arXiv:2012.00407},
  2020.

\bibitem{wu2018survey}
Y.~{Wu}, C.~{Xiao}, Z.~{Ding}, X.~{Gao}, and S.~{Jin}, ``A survey on {MIMO}
  transmission with finite input signals: Technical challenges, advances, and
  future trends,'' \emph{Proceedings of the IEEE}, vol. 106, no.~10, pp.
  1779--1833, 2018.

\bibitem{verdu2002spectral}
S.~Verdu, ``Spectral efficiency in the wideband regime,'' \emph{{IEEE} Trans.
  Inf. Theory}, vol.~48, no.~6, pp. 1319--1343, Jun 2002.

\bibitem{massey1976all}
J.~L. Massey, ``All signal sets centered about the origin are optimal at low
  energy-to-noise ratios on the awgn channel,'' in \emph{Proc. IEEE Int. Symp.
  Inform. Theory (ISIT)}, 1976, pp. 80--81.

\bibitem{lapidoth2002fading}
A.~{Lapidoth} and S.~{Shamai}, ``Fading channels: how perfect need "perfect
  side information" be?'' \emph{{IEEE} Trans. Inf. Theory}, vol.~48, no.~5, pp.
  1118--1134, 2002.

\bibitem{caire2004suboptimality}
G.~Caire, D.~Tuninetti, and S.~Verdu, ``Suboptimality of {TDMA} in the
  low-power regime,'' \emph{{IEEE} Trans. Inf. Theory}, vol.~50, no.~4, pp.
  608--620, April 2004.

\bibitem{li2020single}
Q.~Li, M.~Wen, and M.~Di~Renzo, ``Single-{RF} {MIMO}: From spatial modulation
  to metasurface-based modulation,'' \emph{arXiv preprint arXiv:2009.00789},
  2020.

\bibitem{el2011network}
A.~El~Gamal and Y.-H. Kim, \emph{Network information theory}.\hskip 1em plus
  0.5em minus 0.4em\relax Cambridge university press, 2011.

\bibitem{motahari2014real}
A.~S. Motahari, S.~Oveis-Gharan, M.~A. Maddah-Ali, and A.~K. Khandani, ``Real
  interference alignment: Exploiting the potential of single antenna systems,''
  \emph{{IEEE} Trans. Inf. Theory}, vol.~60, no.~8, pp. 4799--4810, Aug. 2014.

\end{thebibliography}

\end{document}